\documentstyle[aps,prl,twocolumn]{revtex}

\begin{document}

\title{ {\rm\small\hfill submitted to PRL, 7-10-96}\\
Reconstructions and faceting of H covered Al(111)}

\author{Roland Stumpf}
\address{Sandia National Laboratories, Albuquerque, NM~87\,185-1413, USA}
\date{\today}

\maketitle

\begin{abstract}
  First principles calculations show that H overlayers induce a
honeycomb vacancy reconstruction on Al(111).  Above one ML H
coverage, Al(111) facets into surfaces vicinal to (111), with \{100\}
faceted single and double steps and vacancies on the \{111\} terraces.
These H induced reconstructions are stable because H binds more strongly 
to \{100\} than to \{111\} facets.  Faceting of Al(111)
explains step bunching in H-mediated epitaxy of
Al(111). H-covered Al(100) is stable against faceting.
\end{abstract}


Al is an important material for semiconductor devices. Alloyed with a
few percent of Cu and Si, it is the dominant material for
interconnects\cite{Pramanik95}. Shrinking dimensions in chip design
means that control of the Al deposition is more important. Al is usually
deposited under conditions such that H is co-adsorbed on the growing
film. 
However, H reduces the Al film's quality. 
Al(111) grows layer-by-layer at 150\,K in the absence of H\cite{Polanski90}.
Calculations of Al surface diffusion barriers on flat and stepped
Al(111) predict smooth growth at even lower
temperatures\cite{Stumpf96AlD}. With co-adsorbed H as a by-product of
chemical vapor deposition (CVD), Hinch, Doak and Dubois 
report step bunching on Al(111) films grown at 400-500\,K\cite{Hinch93}.

Since H is present in most commercial thin film deposition processes,
understanding the effects of H in epitaxy is of general interest.  H
does not only affect Al epitaxy. For example, H is an anti-surfactant in
Si\cite{Zhang95,Hornvonhoegen95} and Ni\cite{Walters96} epitaxy, while
on GaAs(100)\cite{Morishita95} and C(100)\cite{Frenklach94} H improves
epitaxial growth. 

Another major issue in H-solid interaction is the very common problem of
H embrittlement. Materials that are usually ductile become brittle once
they absorb H. During H promoted fracture, H covers the newly formed
surface. The energy of this H-covered surface can determine the fracture
plane\cite{Stumpf96AlG}.

To study the effect of H on the growth of, e.\,g., Al films, two surface
properties are important: the mobility of Al on H-covered Al surfaces
and the orientation dependence of the surface energy.  A growing Al
surface is rough if the lateral Al mass transport is hindered. A growing
Al film facets if the surface mobility is high and if the total surface
energy can be reduced by forming facets off the growth
direction\cite{Hornvonhoegen95}. To determine surface mobility or
surface energies the structure of the H-covered surface must be known
first, because H can reconstruct surfaces\cite{Stumpf95bh}. Indeed, H
adsorption reconstructs Al(111). As with Be(0001)\cite{Stumpf95bh}, H
induces a honeycomb vacancy reconstruction on Al(111). In this honeycomb
phase 1\,ML of H atoms adsorbs on tilted bridge sites around the
vacancies (see Fig.\,\ref{struct}\,b).  In this very open structure, Al
diffusion is reduced. The effects of H modified surface diffusion will
be studied elsewhere\cite{Stumpf96hd}. In this Letter we concentrate on
understanding the step bunching observed in CVD of
Al(111)\cite{Hinch93}. Our explanation 
is energetically favorable for Al(111) to facet at high H coverage.  The
facet orientation is \{311\} or \{211\}, i.e. vicinal to Al(111) with
\{100\}-faceted steps (see Fig.\,\ref{struct}\,e+h). H induces vacancies
and double steps on those facets. H-covered Al growing in the (100)
direction does not facet, because H binds more strongly to Al(100) than
to other Al surfaces.

{\em First-principles calculations\quad} The study of H adlayers on Al
surfaces reported here is based on the local density
approximation\cite{Ceperley80}. Electronic wavefunctions are expanded in
plane waves up to a cutoff energy of 12\,Ry. All atoms are represented
by pseudopotentials\cite{Hamann89}. We model the Al surfaces by repeated
slabs of 10-20\,\AA{} thickness in orthorhombic or monoclinic
supercells, separated by at least 6\,\AA{} of vacuum. H is adsorbed on
one side only. The adsorbed H atoms plus the top 4-6\,\AA{} of the Al
slab are relaxed. H adsorption energies are quoted relative to 1\,Ry for
free H.  
Special $k$-points at a
density equivalent to 500--5000 $k$-points in the full Brillouin zone of
a one atom cell are used.  Details of the computational technique are
published\cite{Stumpf96AlD,Stumpf95bh,Stumpf94CPC}.




In earlier
publications\cite{Stumpf96AlD,Stumpf95bh,Stumpf96BeHD} we
have accurately calculated properties of simple metal surfaces and of
H-surface interaction. One example is the prediction of a novel type of
H induced reconstruction, the honeycomb vacancy reconstruction of
Be(0001)\cite{Stumpf95bh}. This reconstruction is now confirmed
quantitatively in experiment\cite{Pohl96}. We estimate that adsorption
energy {\em differences\/} are significant if they are larger than 0.05\,eV.
Temperature and zero point motion effects are not included, however.
Furthermore, it is uncertain how accurate calculated surface energies
are, as surface energies are difficult to measure.  Given these
remaining uncertainties, we consider the success of our model in
explaining a number of experimental findings for H on Al surfaces as the
most important evidence for the validity of our theoretical results.


{\em H on flat  Al(111), Al(100) and Al(110)\quad}
Properties of H adlayers on Al surfaces, like adsorption energy and
coordination, depend on surface orientation.  On the flat Al(111)
surface H prefers threefold fcc sites with an adsorption energy ($E_{\rm
ad}$) of 1.12\,eV at 1/12\,ML coverage (see Fig\,\ref{struct}\,a). The hcp
and bridge sites are 0.08\,eV and the top sites are 0.22\,eV higher in
energy. That the fcc site is the prefered threefold site for
H on Al(111) is significant and is discussed elsewhere\cite{Stumpf96hd}.
At higher coverages the adsorption energy is about 0.05\,eV higher
($E_{\rm ad}=1.17$\,eV at 1/2\,ML and $E_{\rm ad}=1.16$\,eV at
1\,ML). Thus, there is a weak attraction between H atoms adsorbed on
Al(111).

On Al(100), H adsorbs preferentially at bridge sites (see
Fig\,\ref{struct}\,f+g). The adsorption energy of H on Al(100) is higher
than on Al(111), with a maximum of 1.33\,eV at 1\,ML.  Below 1\,ML, H
adatoms attract each other ($E_{\rm ad} = 1.22$\,eV at 1/9\,ML). Above
1\,ML the H-H interaction is repulsive ($E_{\rm ad} = 1.26$\,eV at
1.5\,ML and $E_{\rm ad} = 1.09$\,eV at 2\,ML). 
On Al(110), the H adsorption site is coverage dependent.
Up to 1\,ML, H prefers the top site ($E_{\rm ad} =
1.19$\,eV at 1/6\,ML, and $E_{\rm ad} = 1.17$\,eV at 1\,ML). At 2\,ML H
covers the bridge sites of first and second layer ($E_{\rm ad} =
1.17$\,eV).  Thus the H-H interaction is weak.  


{\em H at steps on Al(111)\quad}
It seems that adsorption of H on Al(111), Al(100), and
Al(110) is quite different and unrelated. However, a very close relation
can be established by considering the adsorption of H at close packed
steps on Al(111). There are two types of close packed steps on fcc(111)
surfaces.  One is a \{111\}-microfacet (called B-step), where
the step atoms have the same nearest neighbor environment as
surface atoms on Al(110).  The other step is a
\{100\}-microfacet (called A-step). 


At B-steps H binds as on Al(110) at 1\,ML coverage and
below\cite{stepped}. The adsorption site is the top site, the adsorption
energy, 1.19\,eV, is nearly identical, and the H-H interaction energy is
below 0.01\,eV. At A-steps H binds at bridge sites. The H-H interaction
is attractive with a maximum adsorption energy of 1.35\,eV for one H
adsorbate per step atom. Adsorption site and energy for H at A-steps is
like for H on Al(100).

The fact that H prefers the A-step over the
B-step by as much as 0.16\,eV has important consequences.
One is that H adsorption on Al(111) changes the equilibrium island
shape. Without H the formation energy of A-steps is
higher than that of B-steps (0.248\,eV compared
to 0.232\,eV per step atom)\cite{Stumpf96AlD}. Thus Al islands on clean
Al(111) are hexagonal, with longer B- than A-type edges. With
enough H adsorbed to cover all step atoms, A-steps have less
than half the formation energy of B-steps. This makes the
equilibrium island shape triangular with only A-type edges.  A
change of island shape is not only of aesthetic interest, it also has
been connected to a change in the growth morphology\cite{Michely93}.
The second consequence of the strong binding of H to 
A-steps is that H-covered Al(111) transforms into a stepped
surface with A-steps, as elaborated later.


{\em H adsorbed on Al adatoms\quad}
Up to now we have discussed the adsorption of H on flat and stepped Al
surfaces. Earlier studies of H adsorption on Be(0001) and Be(10$\bar1$0)
show that H can cause vacancy reconstructions of a simple metal
surface\cite{Stumpf95bh,Stumpf96hd,Stumpf96BeHD}. One experimental
signature of the H induced reconstructions of Be(0001) is the high
frequency H vibrational spectrum, which indicates strong bonds of the H
to the reconstructed substrate\cite{Stumpf95bh,Ray90}. Similar modes are
found for H on Al surfaces\cite{Kondoh93}. However, the existing
experimental results give no further hint concerning the structure of
those possible reconstructions. To find a stable
H-induced reconstruction with theoretical tools, a large number of
geometries have to be studied.  To speed up the search 
we identify stable H-Al units of which we then
compose geometries to study.  This also gives insight into H-surface
bonding transferable to other H surface systems. 

The driving force for H-induced reconstructions on simple metal surfaces
is the desire of the H adsorbate to form strong (covalent) bonds to the
substrate.  These bonds are stronger if the metal surface atoms have a
low coordination.  Thus H adsorption lowers the energy of steps, adatoms
and vacancies\cite{Stumpf95bh,Stumpf96BeHD}. In fact, adatoms or
vacancies might even become elements of the most stable surface
geometry\cite{Stumpf95bh}.


As on Be(0001), H binds strongly on top of Al adatoms on all three Al
low index surfaces\cite{Stumpf96BeHD}. The adsorption energy of H on top
of a nearly isolated Al adatom is 1.64\,eV on Al(111), 1.74\,eV on
Al(100) and 1.41\,eV on Al(110).  Thus H is more stable on the adatom
than in the most stable phase on the flat surface.  The energy
difference is 0.47\,eV on Al(111), 0.41\,eV on Al(100), and 0.22\,eV on
Al(110).  On Al(111) even a second H, adsorbed close to the adatom
and on top of a neighboring surface atom, is stable.  The total
energy gain for 2~H atoms adsorbed at the adatom, compared to flat
Al(111), is 0.78\,eV.  The prefered binding of H to adatoms reduces the
adatom formation energy, i.e.\,the energy to remove an Al atom from a
bulk site and put it on the surface. Without H, this energy is 1.05\,eV
on Al(111), 0.56\,eV on Al(100), and 0.29\,eV on Al(110) at low
coverage. With H, the formation energies are reduced to about 0.27\,eV,
0.15\,eV, and 0.07\,eV. These values depend slightly on the H
coverage. On Al(111) the formation energy increases by 0.1\,eV if the
honeycomb phase (see Fig.\,\ref{struct}\,b) is the H reservoir.

Even though Al adatoms are more stable with H co-adsorbed than
without H, we do not find any surface structure involving adatoms that
has a lower energy than the flat surface at the same H coverage.  However,
the equilibrium Al+H concentration should be high on a H-covered Al
surface.  Assuming a room temperature Boltzmann distribution ($kT =
0.025$\,eV), we estimate Al+H concentrations on H-covered Al(100) and
Al(110) of 1/400\,ML and above\cite{AlHdiff}. The Al+H ad-dimer could
be the precursor to Al hydride desorption. This would explain the
observation that up to 10\,\% of the H adsorbed on Al(100) and Al(110)
desorbs as Al hydride\cite{Winkler91}.


{\em H-induced honeycomb reconstruction on Al(111)\quad}
The most stable phase of H-covered Al(111) is the
$\sqrt{3}\!\times\!\sqrt{3}$R$30^\circ$ honeycomb reconstruction.  In
the honeycomb phase, every third Al(111) surface atom is removed and H
atoms cover the bridge sites connecting the remaining atoms (see
Fig.\ref{struct}\,b).  The H adsorbates tilt by 26$^\circ$ in the direction
of the closest fcc site of the ideal surface. The H adsorption energy in
the honeycomb phase ($E_{\rm H\,ad}^{\sqrt3}$) is 1.21\,eV per H,
0.05\,eV higher than for 1\,ML of H on the flat surface.  $E_{\rm
H\,ad}^{\sqrt3}$ is defined as
\begin{equation}
E_{\rm H\,ad}^{\sqrt3} = - 1/3 (E^{\sqrt3} - 3 E^{\rm H} - 
E^{1\times1}_{\rm 5lay} - 2 E^{1\times1}_{\rm 6lay})\;,
\end{equation}
where $E^{\sqrt3}$ is the total energy per unit cell of a $5 + 2/3$
layer Al(111) slab that has a honeycomb array of vacancies on one side.
$E^{\rm H}$ is the energy of a free H atom (1\,Ry), $E^{1\times1}_{\rm 5lay}$ is the energy per unit cell of a 5~layer Al(111) slab
and $E^{1\times1}_{\rm 6lay}$ the energy of a 6~layer slab. This
definition minimizes the errors introduced by finite slab thickness.
H adsorption energies in other structures that involve removing Al atoms
from the surface are defined similarly. 

What makes the honeycomb phase so stable\,? The main reason is that all
H atoms are coordinated like H atoms
adsorbed at A-steps on Al(111) (see Fig.\ref{struct}\,b+h).  The H
adsorption energy at A-steps is 1.35\,eV, 0.19\,eV higher than on
flat Al(111). In a simplistic model, the energy gain per
$\sqrt{3}\!\times\!\sqrt{3}$ unit cell of having more stable H
adsorbates would be 3$\times$0.19\,eV=\,0.57\,eV. This
more than compensates for the vacancy formation energy in a honeycomb
array on Al(111), which is 0.47\,eV. 


Besides the honeycomb phase there are H-induced missing and added row
reconstructions\cite{Stumpf95bh} at 1\,ML H-coverage which are only
0.04\,eV to 0.07\,eV per surface atom higher in energy.
These small energy differences and the small
adatom formation energy should cause the H-covered Al(111) surface to be
disorded at room temperature. This could be the reason why only
1$\times$1 patterns are detected in diffraction experiments. An
alternate reason for the observed 1$\times$1 patterns could be that the
formation of the vacancy reconstructions is kinetically hindered.
However, on Be(0001) the honeycomb structure already forms at
100\,K\cite{Stumpf96BeHD,Pohl96}. Since Be(0001) and Al(111) have
similar surface diffusion barriers\cite{Stumpf96AlD,Stumpf96BeHD}, we
expect that the vacancy reconstructions of H-covered Al(111) are
activated at room temperature.

The fact that H reconstructs Al(111) and that it can even form a surface
hydride at 2\,ML coverage (see Fig.\,\ref{struct}\,c) might be the cause
for the unusual H desorption from Al(111): About half of the H desorbs
as Al hydride and the H$_2$ desorption rate is independent of
coverage\cite{Winkler91}. Details are presented
elsewhere\cite{Stumpf96hd}.

{\em H on surfaces vicinal to Al(111)\quad} To find stable
reconstructions of H-covered Al(111) it is important to consider
surfaces vicinal to Al(111), especially those which have the
energetically favorable A-steps.  Furthermore, since H prefers 
(100) to (111), let us concentrate on vicinal surfaces
with small (111) terraces, i.e., Al(311) and Al(211). (111) terraces
on Al(311) are two rows wide, on Al(211) they are three rows wide (see
Fig.\,\ref{struct}\,e+h).

As on Al(111), it is favorable at high H coverages to form
vacancies on the (111) terraces. These H surrounded
vacancies are most stable if they are in the center of (111) terraces
that are three rows wide.  On ideal Al(311) the (111) terraces are too
short, thus double steps are formed.  The reconstructed (211) and (311)
surfaces are very open and provide for a high concentration of favorable
bridge sites for H adsorption. Nearly all H atoms are adsorbed on
\{100\} facets or at A-steps. The H adsorption energy is
between the values for the honeycomb reconstruction and the flat
Al(100).  The H coverage per area in those (311) and (211) phases,
depicted in Fig.\,\ref{struct}\,e+h), is about 50\,\% higher than ML coverage
on Al(111) or Al(100). Thus the (311) and (211) phases will be
especially stable if the surface H concentration is high.
Incidentially, the H saturation coverage measured on Al(111),
0.19\,\AA$^{-2}$\cite{Winkler91}, coincides with the coverage in the low
coverage phase of the two (311) phases in Fig.\,\ref{struct}. We
therefore suggest that in the measurement of the saturation H coverage on 
Al(111) the surface was locally reconstructed with \{311\} and
possibly \{211\} facets.

{\em Faceting of H-covered Al(111)\quad} The main goal of this Letter is
to explain why H covered Al(111) does not grow smoothly at 400\,K but
forms step bunches and is rough instead\cite{Hinch93}.
We propose that these step bunches are \{311\}- and \{211\}-like
facets which minimize the total surface energy of H covered Al(111).
To compare surface energies of H-covered surfaces we introduce a
construct, the H chemical potential $\mu_{\rm H}$. $\mu_{\rm H}$ is the
energy of H in a fictitious infinite reservoir.  Experimentally an
increase in $\mu_{\rm H}$ corresponds to an increase in H exposure or
coverage.  At large negative values of $\mu_{\rm H}$, H is expensive and
no H adsorbs on the surface.  Without H, the surface energy is $E_{\rm
surf}^0$.  If $\mu_{\rm H}$ is larger than the energy ($- E_{\rm
H\,ad}^1$) of H in the most stable phase (phase 1), then the surface is
covered with a H coverage $c_1$ and the surface energy is
\begin{equation}
E_{\rm surf}^1 (\mu_{\rm H}) =
  E_{\rm surf}^0 - c_1 (\mu_{\rm H} + E_{\rm H\,ad}^1)\,.
\end{equation}
With increasing $\mu_{\rm H}$ higher coverage phases (2, 3, \ldots) 
with smaller $E_{\rm H\,ad}^{2,3,\ldots}$ are formed. 
In principle transitions between different H phases can be used 
experimentally to correlate, e.\,g., H exposure and $\mu_{\rm H}$.

Fig.\,\ref{ehsurf}\,a) shows the surface energies of five Al surfaces 
as a function of $\mu_{\rm H}$. Without H, i.e.\,at small $\mu_{\rm H}$,  
the Al(111) surface is the
most stable. With increasing $\mu_{\rm H}$, the Al(100) surface covered
by 1\,ML H becomes the most stable surface. At even higher $\mu_{\rm H}$,
the reconstructed H-covered Al(311) and Al(211) surfaces take over. 

The surface energies plotted in Fig.\,\ref{ehsurf}\,a) are relevant to
predicting the equilibrium crystal shape of H-covered Al. More
importantly, these surface energies also determine the prefered crack
plane in H induced brittle fracture, since the surface created in H
induced brittle fracture is H-covered.  We speculate that, as in Ga
embrittlement of Al\cite{Stumpf96AlG}, the prefered fracture plane is
the (100) plane. However, further studies are
necessary\cite{Stumpf96hd}.

Fig.\,\ref{ehsurf}\,b) shows the surface energies relevant for faceting
of the Al(111) surface. Except for Al(111), all surface energies are
corrected for the extra surface area that would be created if the
Al(111) surface were to facet. For example, faceting into Al(100)
creates 73\,\% extra surface, which usually costs energy.  Creating
\{311\} or \{211\} facets increases the surface area only little (15\,\%
and 6\,\%).  Therefore, already at a H potential that slightly exceeds
that necessary to adsorb H on the Al(111) surface (in the honeycomb
phase), H-covered Al(111) is unstable against forming trigonal pyramids
with \{311\} or \{211\} faces.  These pyramids might be identical with
what HDD describe as step bunching at a growth temperature of
400\,K\cite{Hinch93}. Step bunching is less pronounced at
500\,K\cite{Hinch93}. The major reason is that at 500\,K the H
desorption rate is much higher than at 400\,K\cite{Kondoh93,Winkler91}.
Thus the H concentration on the surface and therefore $\mu_{\rm H}$, is
reduced.  Thus \{311\} and \{211\} faces are energetically
less favorable during CVD at 500\,K than at 400\,K.

{\em Conclusions\quad}
In this Letter we give a consistent explanation why, in the
presence of H,  Al(111) does not grow smoothly but is rough and forms
facets\cite{Hinch93}. It is rough because H induces a honeycomb vacancy
reconstruction with other reconstructions close in energy. Additionally,
at higher H coverage, and if kinetics allow, H-covered Al(111)
transforms into reconstructed \{311\} or \{211\} facets because this
minimizes the total surface energy. Al(100) is stable against faceting.
This makes the (100) orientation a better candidate for smooth epitaxial
growth. 

The properties of the different H phases proposed here are compatible with
or even suggest an explanation of measurements of the H saturation
coverage, H$_2$ and Al hydride desorption and the H vibrational
frequencies. Clearly, the details have still to be filled in. Also, there
are experimental hints that not only Al(111) but also H-covered Al(100)
and Al(110) reconstruct\cite{Kondoh93}, which deserves further investigation.

{\em Acknowledgments:\quad} Thanks to Peter J. Feibelman and the U.S.
Department of Energy (contract DE-AC04-94AL8500) for their support.

\begin{figure}
\caption{H adsorption energy (eV) for H on five different Al surfaces as
a function of coverage (1/\AA$^2$) and selected diagrams of the atomic
structure.} 
\label{struct}
\end{figure}

\begin{figure}
\caption{Surface energy (meV/\AA$^2$) of five Al surfaces as a function of
the H chemical potential $\mu_{\rm H}$ (eV). To the right the surface
energies are corrected for the additional surface created if Al(111)
facets into a different surface.}
\label{ehsurf}
\end{figure}

\clearpage
\vspace*{5in}\includegraphics{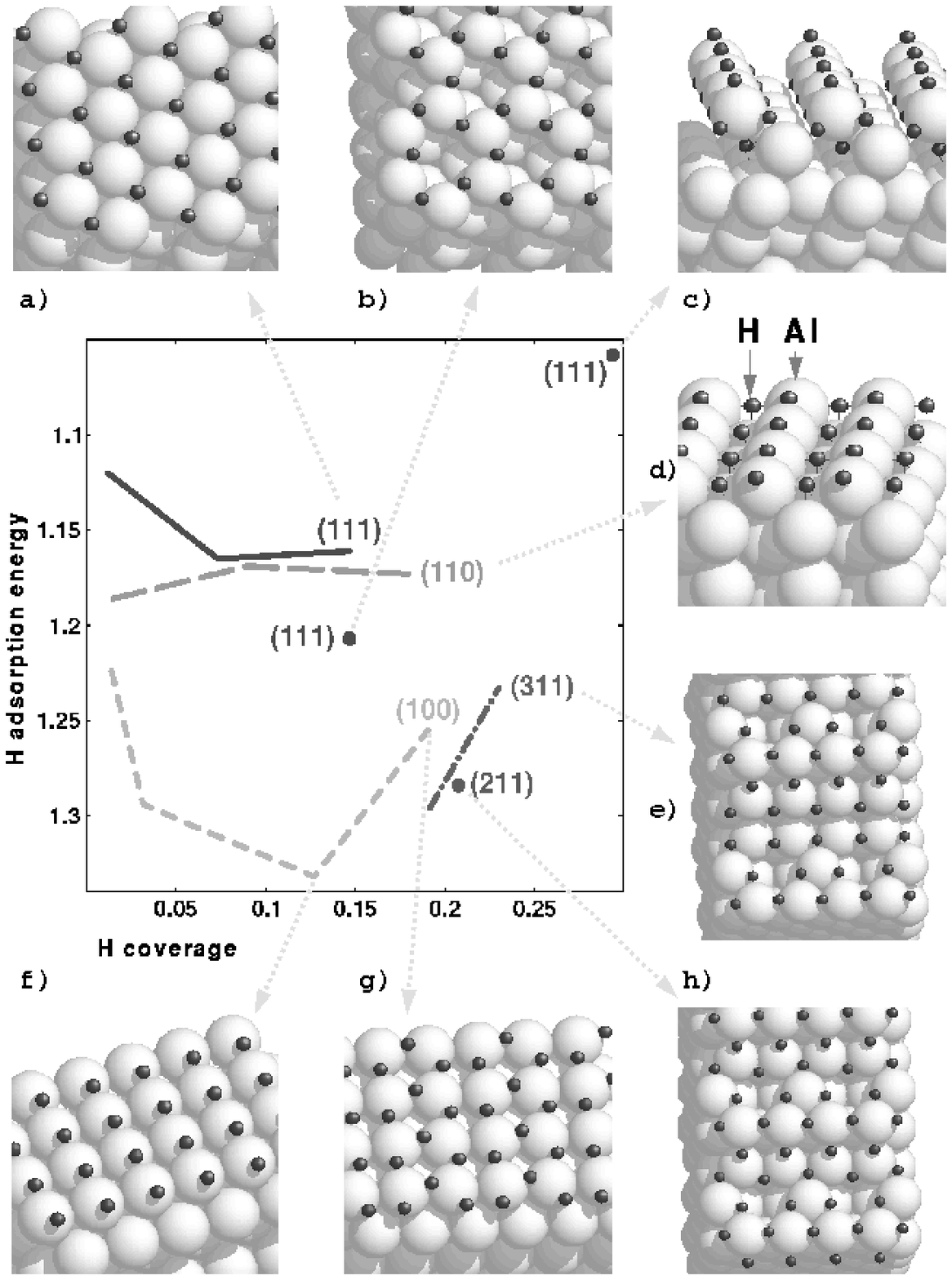}%
\clearpage 
\ \vspace*{5in}\\
\hspace*{1in}\includegraphics{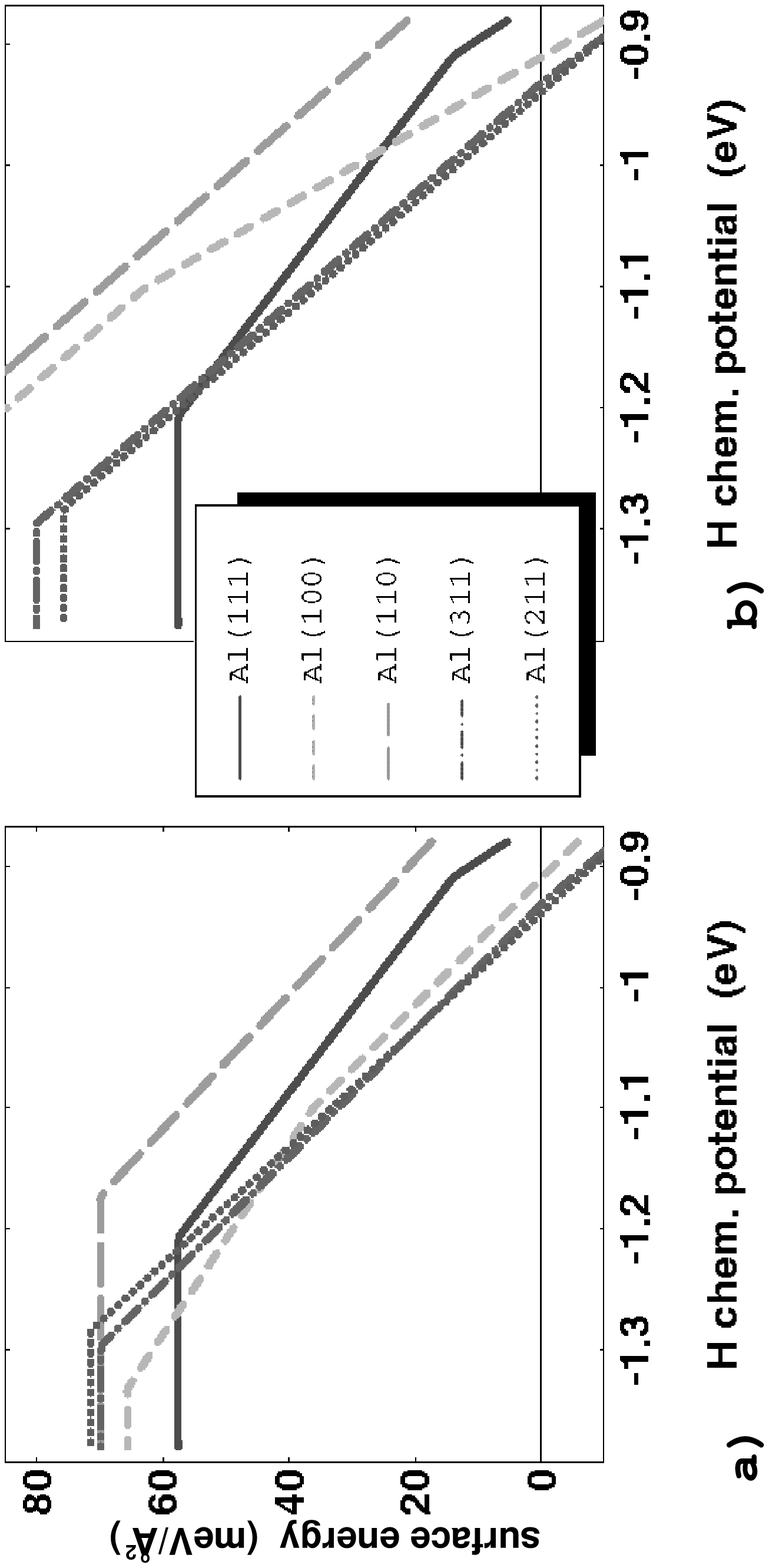}%


\begin{references}
\bibitem{stepped} To study \{111\} facetted steps, we calculate Al(221),
a surface vicinal to Al(111) with Al(111) terraces three atomic rows
wide.  For \{100\} facetted steps we use Al(322), which has (111)
terraces four rows wide.
\bibitem{AlHdiff} Additionally to increasing the concentration of
adatoms, H adsorption also affects the diffusion barrier of the Al(+H)
adatom. Both effects are important for the surface mobility and will be
studied elsewhere\cite{Stumpf96hd}.
\end{references}
\end{document}